\documentclass[twocolumn,aps,prl,amsfonts]{revtex4-1}
\usepackage{amsmath}
\usepackage{graphicx}
\usepackage{bm}
\usepackage{array}
\usepackage{dcolumn}
\usepackage{float}
\usepackage{epstopdf}
\usepackage{lineno}

\newcommand{\up}{\ensuremath{\left|\uparrow\right\rangle}}
\newcommand{\down}{\ensuremath{\left|\downarrow\right\rangle}}

\newcommand{\ket}[1]{\ensuremath{\left|#1\right\rangle}}

\usepackage{braket,mleftright}
\begin{document}

\def\e{\epsilon}
\def\d{\downarrow}
\def\u{\uparrow}
\def\e{\mathcal{E}}
\def\ba{\begin{eqnarray}}
\def\ea{\end{eqnarray}}
\def\beq{\begin{equation}}
\def\eeq{\end{equation}}

\title{Demonstration of two-atom entanglement with ultrafast optical pulses}
\date{\today}
\author{J. D. Wong-Campos}
\email[jwongcam@umd.edu]{}
\author{S. A. Moses}
\author{K. G. Johnson}
\author{ C. Monroe}
\affiliation{Joint Quantum Institute, Joint Center for Quantum Information and Computer Science, and Department of Physics, University of Maryland, College Park, MD 20742} 

\date{\today}

\begin{abstract}

We demonstrate quantum entanglement of two trapped atomic ion qubits using a sequence of ultrafast laser pulses. 
Unlike previous demonstrations of entanglement mediated by the Coulomb interaction, this scheme does not require confinement to the Lamb-Dicke regime and can be less sensitive to ambient noise due to its speed.
To elucidate the physics of an ultrafast phase gate, we generate a high entanglement rate using just 10 pulses, each of $\sim 20$ ps duration, and demonstrate an entangled Bell-state with $(76 \pm 1)$\% fidelity.  These results pave the way for entanglement operations within a large collection of qubits by exciting only local modes of motion.

\end{abstract}

\maketitle

Trapped atomic ions are regarded as one of the most mature and promising platforms for quantum information processing \cite{WinelandBlatt2008,MonroeKim2013}, exhibiting unrivaled coherence properties \cite{Fisk1997}, near-perfect qubit detection efficiency \cite{OxfordDetection2008}, and high-fidelity entangling gates \cite{Lucas1,NISThighfidelity}.  Entangling operations between multiple ions in a chain typically rely on qubit state-dependent forces that modulate their Coulomb-coupled normal modes of motion \cite{CiracZoller, Molmer2, WinelandBlatt2008}. 
However, scaling these operations to large qubit numbers in a single chain must account for the increasing complexity of the normal mode spectrum, and can result in a gate slowdown \cite{InnsbruckShor} or added complexity of the control forces \cite{Shantanu}.

Here we investigate the fundamental entangling operation of a different scaling approach that uses impulsive optical forces \cite{GarciaRipoll,FastDuan,MizrahiPRL, MizrahiAPB, Kielpinski1, Palmero2017}.  These ultrafast qubit state-dependent kicks occur much faster than the normal mode frequencies of motion and thus can couple through local modes of motion without perturbing spectator trapped ion qubits.  This ultrafast approach has the added benefit of being less sensitive to relatively slow noise, and is also insensitive to the ions' thermal motion since it is effectively frozen during the interaction.  Unlike other Coulomb-based gates between ions, ultrafast entanglement operations do not require confinement to within the Lamb-Dicke regime, where the motional extent of the ions is smaller than the optical wavelength associated with the force.  In this Letter, we show a proof-of-principle demonstration of entanglement between two trapped ion qubits by applying a sequence of ten ultrafast laser pulses and directly show the insensitivity to initial thermal motion outside the Lamb-Dicke regime \cite{Thermometry}.

In the experiment, we confine two $^{171}$Yb$^{+}$ atomic ions along the axis of a linear rf (Paul) ion trap \cite {Canlock}.  We apply impulsive forces along one of the transverse principal axes of harmonic motion, coupling to both the in-phase center-of-mass mode at frequency $\omega_C/2\pi = 1.267$ MHz and the out-of-phase relative mode at frequency $\omega_R/2\pi = 1.170$ MHz. The qubit is defined by the ground-state hyperfine levels $|F=0,m_F=0\rangle \equiv \down$ and $|F=1,m_F=0\rangle \equiv \up$ of the $^2 S_{1/2}$ manifold, separated by $\omega_{\text{0}}/2\pi=12.64$ GHz.  The ions are Doppler cooled 
on the $^2 S_{1/2}\Leftrightarrow^2P_{1/2}$ transition at a wavelength of 369.5 nm ($\Gamma/2 \pi\sim 20$ MHz), with both COM and relative modes cooled to an average thermal vibrational population of $\bar{n} \sim 10$.  Qubit state initialization and detection is performed by optical pumping and state-dependent resonance fluorescence on the same transition with fidelities greater than $99\%$ \cite{Olmschenk2007}. Fluorescence is imaged by a 0.6 NA lens with 500x magnification \cite{Imaging}, allowing individual qubit state detection with two separated photomultiplier tubes (PMTs). 

\begin{figure*}[ht]
\includegraphics{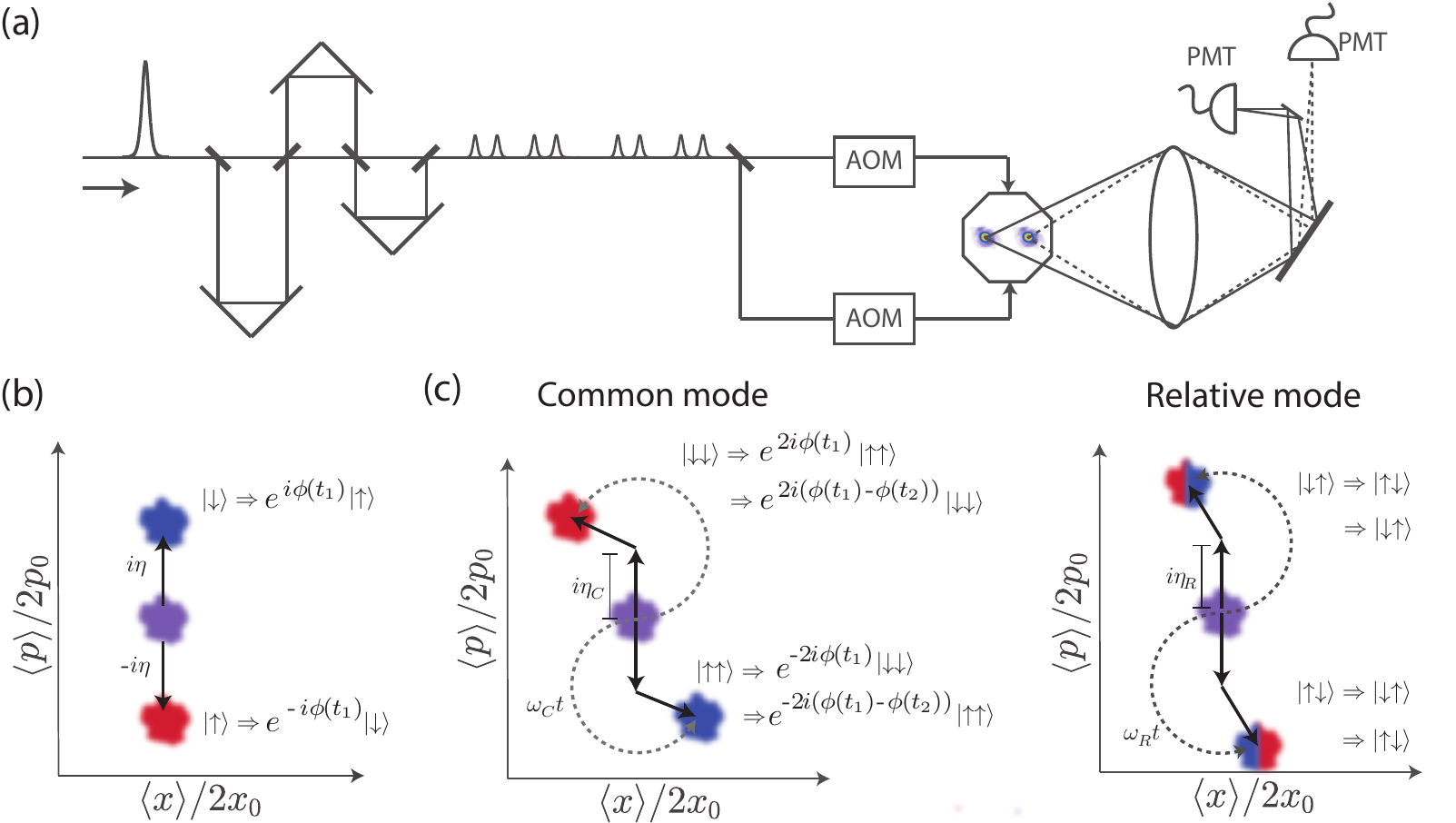}
\caption{(a) Experimental schematic. A single pulse from a mode-locked 355 nm laser is divided into 8 sub-pulses by three sequential optical delay stages. The shaped pulse is then split into two paths directed through independent AOMs, used to make the interaction direction-dependent. The pulses overlap in space and time at the position of the ions in a counterpropagating lin$\perp$lin polarization configuration that produces an SDK \cite{WesPRL}. Following gate operations, the ion qubits are measured by collecting state-dependent fluorescence from the two ions on respective PMTs when resonant lasers are applied (not shown). 
(b) Phase space evolution for a single SDK on a single ion.  The SDK displaces the momentum of an initial motional state (in purple) by $\pm2p_0\eta$ in phase space, correlated with the internal spin raising/lowering operator $\sigma_{\pm}$ in the ion. (c) Phase space evolution for two collective modes of motion under a sequence comprised of an SDK at time $t_1$, free evolution, and a second SDK at time $t_2$, in frames rotating at the respective mode frequencies. Each SDK imparts a phase depending on the magnitude of the momentum kick and the coherent state before the kick (see Eq.~\ref{phieqn}), as well as a laser phase $2 \phi(t) (-2 \phi(t))$ on the $\lvert \d \d \rangle$ ($\lvert \u \u \rangle$) states (see Eq.~\ref{gammaeqn}).  In addition, the motional states acquire a phase from the free evolution (see Eq.~\ref{displacement}).  Because the evolution is depicted in the rotating frame, the direction of an SDK depends on the time elapsed since the previous SDK.
}\label{fig:Setup}
\end{figure*}

The impulsive forces are provided by $\tau \sim 20$ ps pulses from a mode-locked laser with center wavelength $2\pi/k=355$ nm and repetition rate $f_{\text{rep}}= 81.42$ MHz that drives stimulated Raman transitions between the qubit levels \cite{WesPRL}.  As shown in Fig.~\ref{fig:Setup}a, after picking single pulses with an electrooptic Pockels Cell, we shape each pulse using a sequence of three delay stages to divide each single pulse into eight sub-pulses in order to achieve the desired spin-motion dependence \cite{MizrahiPRL}. The pulse train is split into two arms with orthogonal linear polarizations and directed onto the ions in a counter-propagating geometry along the transverse direction of motion.  Each arm includes an acousto-optic modulator (AOM) that shifts the center frequency of each beam with opposite sign. The net frequency difference between the two arms is set to $\omega_A= \omega_{\text{AOM1}} + \omega_{\text{AOM2}} = 2\pi \times  468.73$ MHz.  The delays between the eight sub-pulses are set in concert with the frequency offset $\omega_A$ in order to produce a qubit state-dependent kick (SDK) \cite{MizrahiAPB}.  

For a single trapped ion, the ideal evolution operator from an SDK applied at time $t$ is given by
$e^{i\phi(t)}\hat{\mathcal{D}}(i\eta)\hat{\sigma}_{+} + e^{-i\phi(t)}\hat{\mathcal{D}}(-i\eta)\hat{\sigma}_{-}$.
The phase $\phi(t)=\omega_A t + \phi_L$ is related to the AOM frequency and the absolute optical phase $\phi_L$ of the driving laser, assumed to be common mode for the two beam paths and constant during the interaction.  The raising and lowering operators $\hat{\sigma}_{\pm}$ act on the qubit, and the displacement operator $\hat{\mathcal{D}}(\pm i\eta)$ acts on the motional state of the ion along the axis of transverse motion, translating the momentum in phase space by $\Delta p = \pm\hbar(\Delta k) = \pm 2 p_0 \eta$.  Here $\Delta k = 2k$ is the wavevector difference between the counter-propagating beams and $p_0 = \sqrt{m\hbar\omega/2}$ is the zero-point momentum spread of harmonic motion at frequency $\omega$ for an ion of mass $m$ ($x_0=\hbar/2p_0$ is the zero-point position spread). The Lamb-Dicke parameter $\eta = \hbar \Delta k /(2 p_0) \approx 0.17$ thus parametrizes the momentum kick in natural units. In contrast to conventional forces applied in the resolved sideband regime \cite{newtestament}, the impulsive SDK is about three hundred times faster than the oscillation period and does not rely on confinement to the Lamb-Dicke regime.

The action of an SDK on two ions is given by
\begin{equation}
\begin{aligned}
\hat{U}_{SDK}(t) = &\  e^{2i\phi(t)}\hat{\sigma}_{1+}\hat{\sigma}_{2+}\hat{\mathcal{D}}_C( i\eta_C) +   \\
                   &\ e^{-2i\phi(t)}\hat{\sigma}_{1-}\hat{\sigma}_{2-}\hat{\mathcal{D}}_C(-i\eta_C) +  \\
                   &\ \hat{\sigma}_{1+}\hat{\sigma}_{2-}\hat{\mathcal{D}}_R( i\eta_R) +\hat{\sigma}_{1-}\hat{\sigma}_{2+}\hat{\mathcal{D}}_R(-i\eta_R), 
\end{aligned}
\end{equation}
with spin operators for each ion and displacement operators for each mode.  The Lamb-Dicke parameters for the COM and relative modes are $\eta_C = \sqrt{2}\eta=0.24$ and $\eta_R = \sqrt{2\omega_C/\omega_R}\eta=0.25$. 
Due to their distinct displacement amplitudes and frequencies, the COM and relative modes trace distinct paths in phase space when subjected to a sequence of SDKs interspersed with free evolution. Fig.~\ref{fig:Setup}c shows the trajectories of the two modes in frames rotating at the respective mode frequencies, where the SDK displacement for each mode $m$ has magnitude $\eta_{m}$ along an axis rotated by angle $\omega_{m}t$ with respect to the previous kick after elapsed time $t$.  

\begin{figure*}[ht]
\includegraphics{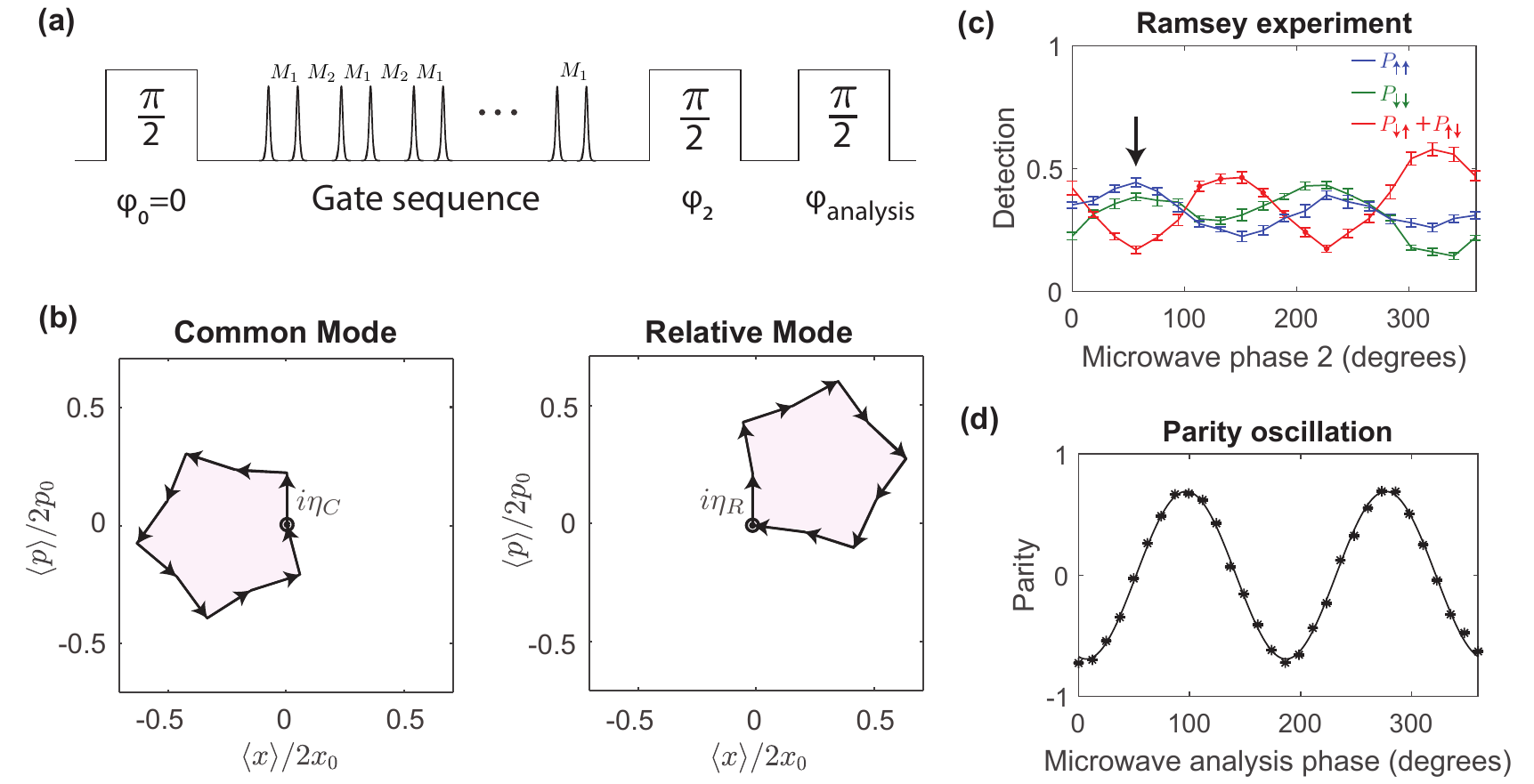}
\caption{(a) The gate sequence is applied within a Ramsey experiment. The entangling gate contains 5 repetitions of a sequence consisting of a single SDK ($b_n=1$), followed by a wait of $M_1$ time steps ($b_n=0$), another SDK ($b_n=1$), and a final wait of $M_2$ time steps. (b) Depiction of the trajectories followed by the COM and relative modes for a fully entangling sequence. They follow opposite circulations, enclose similar areas, and the sum leads to the gate phase. (c) We choose $\phi_2$ such that we maximize $P_{\uparrow \uparrow}$, $P_{\downarrow \downarrow}$ and minimize $P_{\downarrow \uparrow}$, $P_{\uparrow \downarrow}$ (black arrow). Population in $\ket{\uparrow \downarrow}$ and $\ket{\downarrow \uparrow}$ is due mostly to SDK infidelity. (d) The parity oscillation amplitude after choosing the leftmost value of $\phi_2$ in (c) is proportional to $2\rho_{\uparrow \uparrow, \downarrow \downarrow}$, which allows us to compute the fidelity. The best parity oscillation amplitude achieved is 0.69(1), leading to a final gate fidelity of 76(1)\%. 
} 
\label{fig:piover2}
\end{figure*}

A sequence of SDK pulses indexed to uniform time steps of duration $T=1/f_{\text{rep}}$ can be expressed by $N$ displacement indices  $\{b_1,b_2,...,b_N\}$ with $b_n = 1$ corresponding to a kick as described above, $b_n=-1$ corresponding to a kick with reversed beam directions $(\Delta k \rightarrow -\Delta k)$, and $b_n= 0$ corresponding to a wait (no pulse).  This sequence leads to displacements of initial coherent states for each mode $m$ from $|\alpha_0\rangle_m$ to $e^{i\phi_m}|\alpha\rangle_m$, with \cite{GarciaRipoll}
\begin{eqnarray}
\alpha &=& e^{-iN\omega_m T}\left( \alpha_0 + i\sum^N_{n=1} \eta_m b_n e^{in\omega_m T}\right) \label{displacement} \\
\phi_m &=& \mathcal{R}e\left(\alpha_0 \sum^N_{n=1} \eta_m b_n e^{-in\omega_m T}\right) \nonumber \\ 
&& + \sum^N_{n=2}\sum^{n-1}_{j=1}\eta_m^2 b_n b_j \sin[\omega_mT(n-j)].\label{phieqn}
\end{eqnarray}
We design pulse sequences $\{b_{n}\}$ so that the sum in Eq. \ref{displacement} vanishes and both motional phase spaces close.  
Given an even number of pulses, this produces a phase gate described with truth table 
\begin{equation}
\begin{aligned}
\lvert\downarrow\downarrow \rangle &\Rightarrow  \lvert\downarrow\downarrow \rangle e^{i(\Phi_g+\gamma)}   \\
\lvert\downarrow\uparrow \rangle &\Rightarrow \lvert\downarrow\uparrow \rangle \\
\lvert\uparrow\downarrow \rangle &\Rightarrow \lvert\uparrow\downarrow \rangle  \\
\lvert\uparrow\uparrow \rangle &\Rightarrow \lvert\uparrow\uparrow \rangle e^{i(\Phi_g-\gamma)}.
\label{truth}
\end{aligned}
\end{equation}
The nonlinear geometric phase $\Phi_g = \phi_C - \phi_R$ is set to $\pi/2$ for maximum entanglement.  The residual linear phase from the series of optical kicks is 
\begin{equation}\label{gammaeqn}
 \gamma= 2\omega_A T \sum^N_{n=1} (-1)^{\sum_{j=1}^n |b_j|+1} n |b_n|,
\end{equation}
where the alternating signs account for the qubit spin flip after each SDK.  Note that because the number of pulses is even, the net gate evolution is insensitive to the optical phase $\phi_L$, which is assumed to be constant over the course of the gate.

Here, we implement a quantum gate with fast pulses by finding gate sequences with the least number of SDKs, without reversing the beam directions (restricting $b_n = 0$ or $1$).  For $N_p$ individual pulses separated in time by an integer multiple $M$ of the laser pulse period $T$, the condition for closing phase spaces is similar to the tracing of a regular polygon in the complex plane.
We achieve the largest nonlinear gate phase for a given number of pulses by driving the COM and relative modes in opposite directions in phase space so that $\phi_R \approx -\phi_C$.

\begin{figure}[t!]
\includegraphics{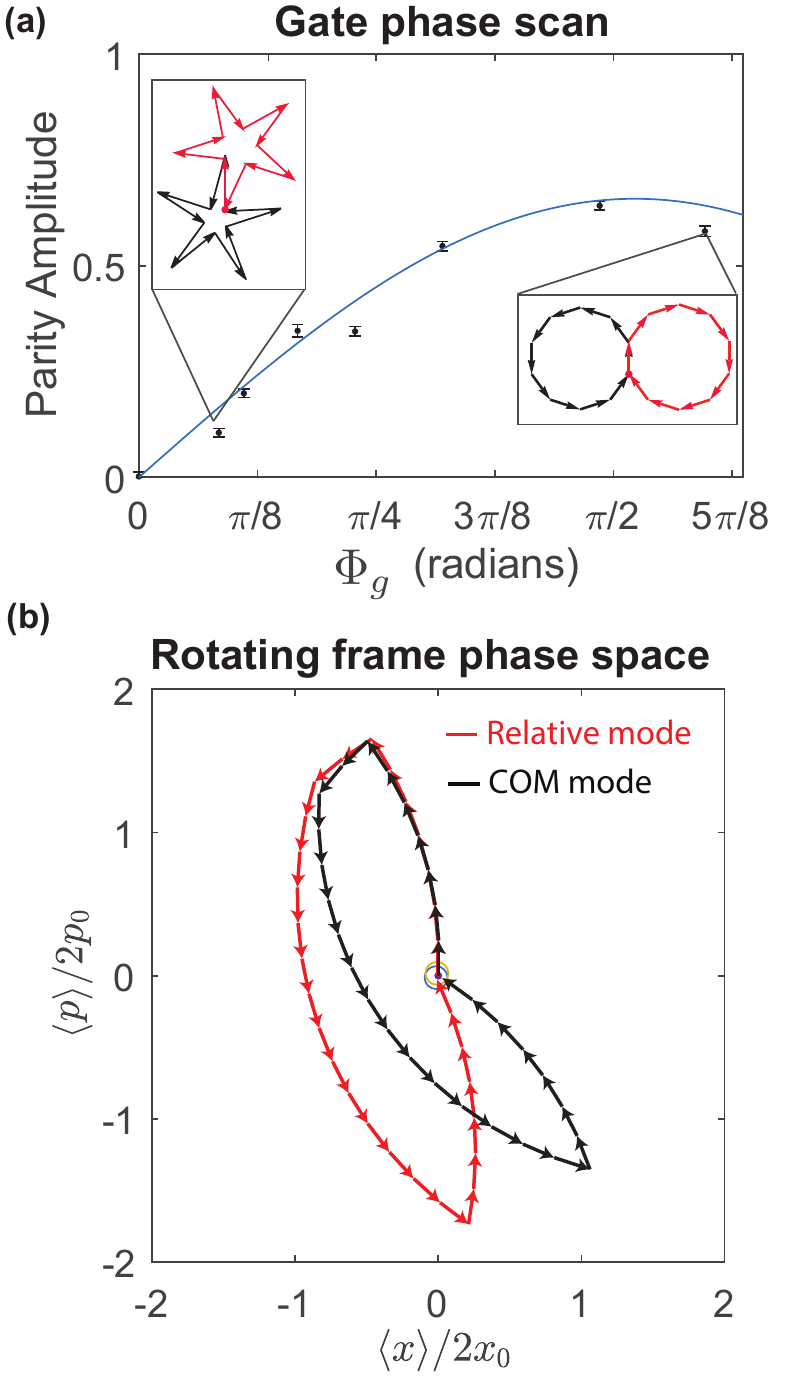}
\caption{(a) Measured parity oscillation amplitude for various values of the gate phase $\Phi_g$, which should be proportional to $\sin \Phi_g$ (solid blue line).  The gate phase is modified by changing the pulse schedule given by the integers $M_1$ and $M_2$ (see main text). The insets show phase space trajectories for the COM (red) and relative (black) modes for $\Phi_g= \pi/1.67$ (right) and $\Phi_g = \pi/11.9$ (left).  The fidelity of the entangled state produced in each case, referenced to the ideal state $\Psi_f(\Phi_g)$ in Eq. \ref {eq:psiF}, is roughly 0.7 for all the measurements. (b) Theoretical pulse sequence for a maximally entangling gate made out of 30 SDKs with a total gate time of 921 ns. 
The modes are kicked consecutively by switching the directions of the beams to create a  ``lever arm" for geometrical phase acquisition.
}
\label{fig:varynandm}
\end{figure}

Using the above trap parameters with $N_p=10$, we find that the phase space trajectories of COM and relative modes trace out regular decagons of opposite circulation for $M=166$, with corresponding gate phase $\Phi_g = \pi/1.67$.  Other values of $\Phi_g$ can be realized by alternating between two different integer multiples of the pulse periods, $M_1$ and $M_2$, such that $M_1+M_2 =2M$. This deforms the trajectories to decagons with two distinct vertex angles (see Figs.~\ref{fig:piover2}b and \ref{fig:varynandm}a), allowing the fine tuning of $\Phi_g$. For $M_1=175$, $M_2=157$, we find $\Phi_g= \pi/2.06$, nearly a fully entangling gate in a total duration of 
$(N_p M-M_2)T = 18.5$ $\mu$s.  There are many more types of pulse solutions with even more complex polygonal trajectories given the delay times between pulses.

We characterize the phase gate by applying the gate operation within a three-pulse Ramsey interferometer on the qubits. We start the sequence by optically pumping the ions to the state $\lvert\downarrow\downarrow \rangle$. A first microwave $\pi/2$-pulse rotates both spins to populate an equal superposition of all 4 basis states. The entangling laser pulse sequence is then applied, which according to the truth table (eq. \ref{truth}) should ideally produce the state
\begin{equation}
\Psi_e=\frac{e^{i\Phi_g}}{2}\left(e^{i\gamma}\lvert\downarrow\downarrow \rangle + e^{-i\gamma}\lvert\uparrow\uparrow \rangle\right)-\frac{1}{2}\left(\lvert\uparrow\downarrow \rangle + \lvert\downarrow\uparrow \rangle\right),
\end{equation} 
where in the above expression we have suppressed the motional state, since both phase spaces should be closed at this point. 

A second $\pi/2$ microwave Ramsey pulse of variable phase with respect to the first pulse is then applied. We choose its phase to ideally create the state
\begin{equation}
\Psi_f = \frac{e^{-i\gamma}}{2} (e^{i\Phi_g}-1)\lvert\downarrow\downarrow \rangle + \frac{e^{i\gamma}}{2} (e^{i\Phi_g}+1) \lvert\uparrow\uparrow \rangle.
\label{eq:psiF}
\end{equation}
We experimentally determine the appropriate phase of the second Ramsey $\pi/2$ pulse by maximizing the populations $P_{\uparrow\uparrow}$ and $P_{\downarrow\downarrow}$ of the even parity states, as shown in the Ramsey fringes of Fig.~\ref{fig:piover2}c.

In order to verify the coherence of the above entangled state, we apply a third $\pi/2$ ``analysis pulse" and measure the parity of the two qubits as a function of this last pulse, as shown in Fig.~\ref{fig:piover2}d. The parity oscillates with twice the period of a single spin, and the contrast $C$ of the oscillation reveals the coherence between the entangled superposition in Eq. \ref{eq:psiF}.  The state fidelity with respect to the ideal Bell state is then
$F=(P_{\uparrow\uparrow}+ P_{\downarrow\downarrow}+C)/2$  \cite{fourparticles}. We measure a Bell state fidelity of $F=76(1)\%$.

As a further validation of our control over various gate sequences, we vary the gate phase $\Phi_g$ by changing the number of pulses $M_1$ and $M_2$ over a wider range.  In Fig.~\ref{fig:varynandm}a we show a measurement of the parity oscillation contrast $C$ for different values of $\Phi_g$. The measured parity oscillation amplitude for each gate sequence agrees well with the expected $\sin \Phi_g$ dependence.
Finally, we note that the linear phase $\gamma$ can be regarded as a constant offset phase in the above data and does not affect the amount of entanglement or its diagnosis. 

The entangling gate presented here is fundamentally different than the M\o lmer-S\o rensen \cite{Molmer2} and Cirac-Zoller \cite{CiracZoller} gates for trapped ions, since individual motional modes are not resolved.  Moreover, the (thermal) motion of the ions occupies a spatial extent of $x_0\sqrt{2\bar{n}+1} \approx 0.8/\Delta k$, outside the Lamb-Dicke regime. 

The gate can be made much faster by dynamically switching the laser beam wavevector difference $\Delta k$ and thus using negative values of $b_n$ in the pulse schedule to close phase spaces more quickly \cite{FastDuan}. 
We have identified potential gate sequences with net gate times shorter than 1 $\mu$s, as shown in Fig.~\ref{fig:varynandm}b, with the same trap parameters used above. Experimentally, fast switching can be accomplished by inserting a second electrooptic Pockels cell after the two AOMs, as previously demonstrated for the generation of large Schr\"{o}dinger cat states \cite{CatKale}.  However, extension of this setup for high-fidelity 2-qubit gates remains challenging because of instabilities in the positioning and polarization of the high power ultraviolet beams traversing the Pockels cell. 
In the future, it may be possible to control infrared optical sources instead of the ultraviolet lasers used here, by frequency-upconverting to the ultraviolet after pulse-shaping/switching, or exploiting a longer-wavelength atomic transition for the SDK.

\section{Acknowledgements}
This work is supported by the U.S. Army Research Office and the NSF Physics Frontier Center at JQI.

\bibliographystyle{prsty}
\bibliography{ReferencesFastGates}

\end{document}